\documentclass[preprint,showpacs,amsmath,amssymb]{revtex4}
\usepackage{amssymb,graphicx}
\usepackage{dcolumn}% Align table columns on decimal point

\def\P3{{\cal P}_t}
\def\J3{{\cal J}}
\def\T3{{\cal T}}

\def\v#1{{\bf#1}}
\def\bra{\langle}
\def\ket{\rangle}

\def\egu{\, =\, }
\def\plus{\, +\,}

\def\cap{\noindent}

\def\hskstm{ {{\hbar^2\v{k}^2}\over {2 m}} }

\def\beq{\begin{equation}}
\def\eeq{\end{equation}}
\def\bar{\begin{array}[b]}
\def\barc{\begin{array}}
\def\bart{\begin{array}[t]}
\def\ear{\end{array}}
\def\le#1{\label{eq:#1}}
\def\re#1{\ref{eq:#1}}

 1  1
 1 
scaled\magstephalf 
 
scaled\magstephalf

\begin{document}
\thispagestyle{empty}
\vspace*{0.5 cm}
%\vspace*{-1.2in}
%\vspace*{0.7 in}
\begin{center}
{\bf Neutron matter at low density and the unitary limit.}
\\
\vspace*{1cm} {\bf M. Baldo and C. Maieron}\\
\vspace*{.3cm}
{\it Dipartimento di Fisica, Universit\`a di Catania}\\
and \\
{\it INFN, Sezione di Catania}\\
{\it Via S. Sofia 64, I-95123, Catania, Italy} \\
\vspace*{.6cm}
%\vspace*{.6cm} {\bf }\\ \vspace*{.3cm}
%{\it INFN, Sezione di Catania}\\
%{\it Dipartimento di Fisica, Universit\`a di Catania}\\
%{\it Corso Italia 57, I-95129, Catania, Italy} \\
\vspace*{1 cm}
\end{center}
{\bf ABSTRACT} \\
Neutron matter at low density is studied within the hole-line expansion.
Calculations are performed in the range of  Fermi momentum $k_F$ between $0.4$
and $0.8$ fm$^{-1}$. It is found that the Equation of State is determined by
the $^1S_0$ channel only, the three-body forces contribution is quite small,
the effect of the single particle potential is negligible and the three
hole-line contribution is below 5\% of the total energy and indeed vanishing
small at the lowest densities. Despite the unitary limit is actually never
reached, the total energy stays very close to one half of the free gas value
throughout the considered density range. A rank one separable representation of
the bare NN interaction, which reproduces the physical scattering length and
effective range, gives results almost indistinguishable from the full Brueckner
G-matrix calculations with a realistic force. The extension of the calculations
below $k_F = 0.4$ fm$^{-1}$ does not indicate any pathological behavior of the
neutron Equation of State. \vskip 0.3 cm
%\noindent
%\vfill\eject
\section{Introduction}
The crust of neutron stars is composed of a solid lattice of nuclei, whose
masses and neutron excess increase as one proceeds from the surface to the
interior \cite{shap}. This is due to the increase of the matter density and of
the corresponding electron density, which shifts the beta equilibrium towards
larger and larger nuclear asymmetry. At a definite density nuclei start to drip
neutrons since their chemical potential turns positive. The inner crust is then
formed by a nuclear lattice permeated by a gas of neutrons. From the drip point
on, the neutron gas density increases, starting in principle from a vanishing
small value, up to the point where nuclei merge and possibly form more
complicated structures and finally a homogeneous matter of neutrons and protons
appears. This is one of the main reasons of the great interest that has been
devoted to the study of the Equation of State (EOS)  of pure neutron matter.
 The low density region is less trivial than one
could expect at a first sight since the neutron-neutron scattering length is
extremely large, about $-18$ fm, due to the well known virtual state in the
$^1S_0$ channel, and therefore even at very low density one cannot assume the
neutrons to be uncorrelated. These considerations have also stimulated a great
interest in the so called unitary limit, i.e. the limit of infinite (negative)
scattering length of a gas of fermions at vanishing small density. A series of
works \cite{carl,boro,bulg1} have been presented in the literature based on
various approximations and a recent Monte Carlo calculation \cite{bulg2} on a
related physical system has shown that the unitary limit can present a quite
complex structure, involving both fermionic and bosonic effective degrees of
freedom, which has still to be elucidated. Variational \cite{panda1} and finite
volume Green's function Monte Carlo calculations  \cite{panda2} for neutron
matter at relatively low density have shown that the EOS, in a definite density
range, can be written as the free gas EOS multiplied by a factor $\xi$, which
turns out to be close to $0.5$. This is actually what one could expect in the
unitary limit regime, since no scale exists in this case, except the Fermi
momentum. Monte-Carlo calculations \cite{carl,boro,bulg1} with schematic forces
in a regime close to the unitary limit have found a factor $\xi \approx 0.44$.
The connection between the variational results and the unitary limit has been
studied in ref. \cite{Pethick} by means of effective theory methods.
\par In this paper we
present results on pure neutron matter EOS based on the hole-line expansion of
Bethe, Brueckner and Goldstone (BBG) \cite{book}, which is particularly suited
for the low density regime. We use a realistic force, as specified below, and
we show that at the lowest density a rank one separable representation of the
neutron-neutron interaction, which incorporates the physical values of the
scattering length and effective range, is able to reproduce accurately the EOS
obtained with the full glory NN interaction. The extension of the calculations
to very low density is also discussed.
\section{The in medium G-matrix}
Since the scattering length $a$ and effective range $r_0$ in the $^1S_0$
channel of the neutron-neutron interaction differ by about a factor 6, there is
no density interval where the unitary limit can be considered strictly valid.
However, in the range  $r_0 < d < |a|$, where $d$ is the average inter-particle
distance, the physical situation should be the ``closest" possible to the
unitary limit.
 For the sake of comparison we first restrict the analysis to the density range
corresponding to $0.4$ fm$^{-1}$ $<  k_F  <$  $0.8$  fm$^{-1}$, which falls in
this range and corresponds to densities between about 1/50 and 1/5 of the
saturation density. As a modern realistic nucleon-nucleon potential we choose
the Argonne v$_{18}$ interaction \cite{wir}. From the three-body force of the
Urbana model, adjusted to reproduce the correct saturation point \cite{bbb}, we
found a contribution which is less than $0.01$ MeV, and therefore we neglect
three-body forces hereafter.
\par The basic quantity in the BBG expansion is the Brueckner G-matrix, which
satisfies the integral equation \beq \bar{rl}
 \bra k_1 k_2 \vert G(\omega) \vert k_3 k_4 \ket\!\!\! &\egu
 \bra k_1 k_2 \vert v \vert k_3 k_4 \ket \plus  \\
 &                  \\
 + \sum_{k'_3 k'_4} \bra k_1 k_2 \vert v \vert k'_3 k'_4 \ket\!\!\!
 &{\left(1 - \Theta_F(k'_3)\right) \left(1 - \Theta_F(k'_4)\right)
  \over \omega - e_{k'_3} - e_{k'_4} }
  \, \bra k'_3 k'_4 \vert G(\omega) \vert k_3 k_4 \ket \ \ .
\ear \le{bruin} \eeq \cap

\begin{figure}[b]
\vspace {-3 cm} \centerline{\includegraphics
[height=240mm,width=180mm]{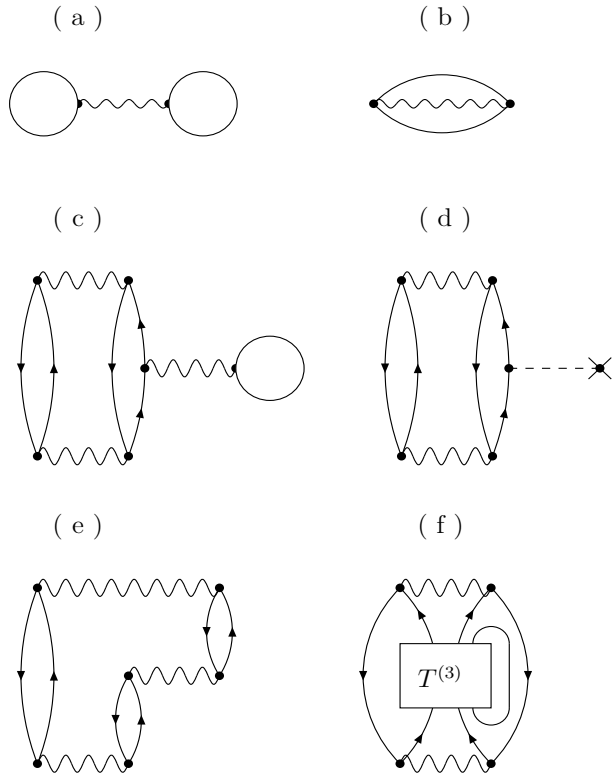}} \vspace {-9.6 cm} \caption{Two and three
hole-line diagrams for the ground state energy in terms of the G-matrix (wiggly
lines).} \label{fig:bhf3}
\end{figure}

The intermediate states are particle states, and this is indicated in Eq.
(\re{bruin}) by the two Pauli projection factors $1 - \Theta_F(k)$, being
$\Theta_F(k)$ the Fermi distribution at zero temperature. The entry energy
$\omega$ is specific of each diagram where the G-matrix appears. At the two
hole-line level of approximation, the diagrams which contribute are the ones
indicated by labels {(a)} and  {(b)} in Fig. \ref{fig:bhf3} (first row), where
the wiggly line indicates the Brueckner G-matrix.

\begin{figure}[]
\centerline{\includegraphics [height=140mm,width=140mm]{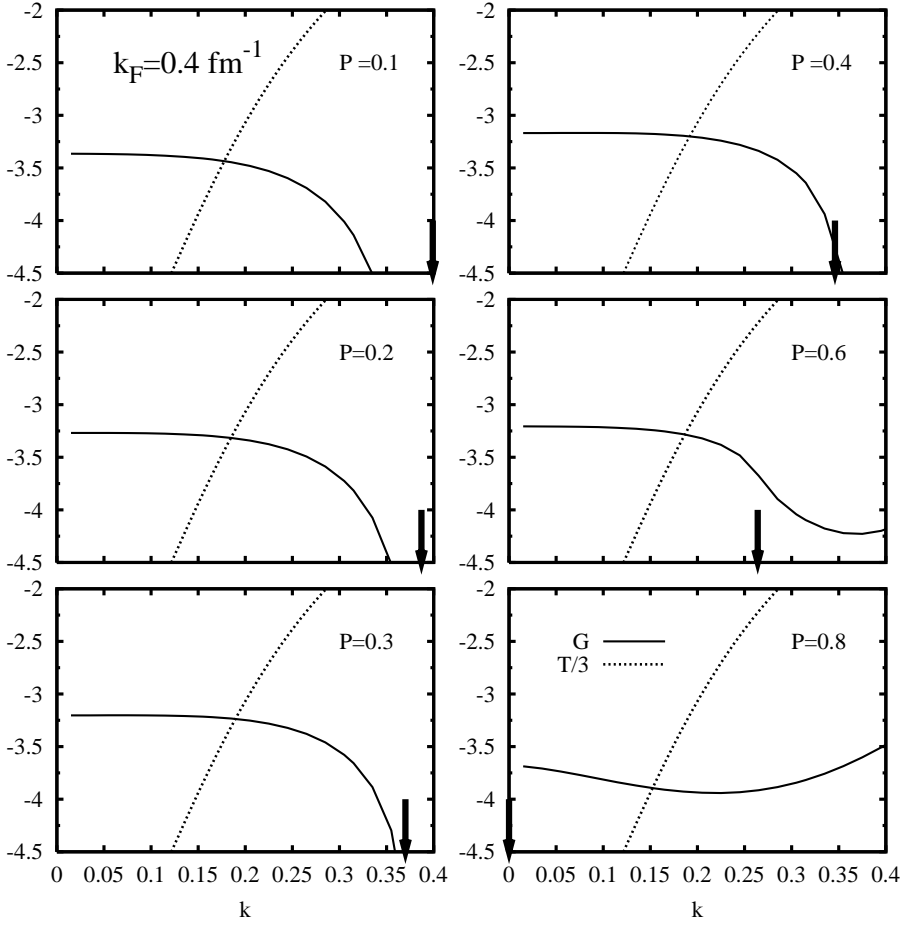}}
 \caption{Comparison between the free T-matrix and the Brueckner G-matrix
 at different total momentum {P} and relative momentum {k}  (in fm$^{-1}$) at the Fermi
 momentum ${k_F  =  0.4}$fm$^{-1}$. The arrows indicate the upper limit of the
 momentum integration needed for
calculating the interaction energy. } \label{fig:pauli}
\end{figure}

They correspond to the Brueckner-Hartree-Fock (BHF) approximation. For these
diagrams the entry energy is just the energy of the two interacting particles.
It is remarkable that even at the relatively low densities we are considering
the inclusion of the Pauli projection in the intermediate states produces the
overwhelming dominant in medium effect. This is illustrated in Fig.
\ref{fig:pauli}, where the diagonal G-matrix in the $^1S_0$ channel is reported
in comparison with the corresponding free T-matrix (divided by 3) at selected
values of the relative momentum $k$ and total momentum $P$ (in fm$^{-1}$) at
the Fermi momentum $k_F = 0.4$ fm$^{-1}$.

Of course, due to Galilei invariance, the free T-matrix is independent of $P$.
For simplicity the free single particle spectrum (kinetic energy) is adopted in
these calculations, but, as we will see, this is not a serious restriction.
Despite the Fermi momentum is quite small, a drastic difference between the two
scattering matrices is apparent, not only in shape but also in absolute value.
The Pauli operator effect is enhanced in this particular channel since the
virtual state is suppressed in the medium, as we will discuss later in detail.
The arrows indicate the upper limit of the momentum integration needed for
calculating the interaction energy. To be noticed is also the pairing
singularity at the Fermi momentum and for small total momentum $P$. This
singularity is integrable and can be handled without numerical problems. This
means of course that we are neglecting the pairing condensation energy, which
is however negligible in this density range. In any case we are going to
compare in a coherent fashion only calculations that neglect the pairing
contribution.
\par
Since s-wave dominates at low density, the in medium modification of the
G-matrix in the $^1S_0$ channel has a profound and essential effect on the EOS.
Indeed, higher partial waves give a negligible contribution to the interaction
energy.
%, as shown in Fig. \ref{fig:waves}, where the potential energy
%contributions in the BHF approximation up to $f$-wave are reported.
In practice the calculation of the EOS  at the BHF level of approximation
reduces to a single channel problem in the considered density range (or below).
The inclusion of higher partial waves would not alter at all our analysis and
conclusions.
\par
In the BBG expansion an auxiliary single particle potential $U(k)$ is
introduced. Then the single particle energy reads
\beq
 e(k)\, =\, \hskstm \, +\, U(k)
\eeq

\noindent and the potential $U(k)$ is determined in a self-consistent way in
terms of the Brueckner G-matrix
\beq
 U(k) \egu \sum_{k' < k_F}
      \bra k k' \vert G(e_{k_1}+e_{k_2}) \vert k  k' \ket \ \ \ ,
\le{auxu}
\eeq
\noindent

The auxiliary potential is essential to get convergence in the BBG expansion
\cite{book,jpg}. However in the low density regime we are considering we found
that its effect on the G-matrix is negligible, at least within the accuracy we
need for the considerations developed in the present work. This does not mean
that we can neglect $U(k)$ altogether, since, as we will see, at the three
hole-line level of approximation its effect through the ``$U$-insertion"
diagram is of some relevance.
\section{The three hole-line contribution and the EOS.}
The BBG expansion relies on the basic idea that the contributions of the
diagrams of the expansion decrease with increasing number of hole-lines which
are included. Despite that BBG is essentially a low density expansion, it has
been found \cite{bal1,bal2} that the convergence is valid up to densities as
high as few times saturation density in symmetric nuclear matter and even
better in neutron matter. It is then likely that at the low densities we are
here considering this convergence should be even faster. This is indeed
confirmed by explicit numerical calculations reported below. The three
hole-line diagrams can be summed up by means of the Bethe-Fadeev equation
\cite{bethe}, which introduces the in medium three-body scattering matrix
$T^{(3)}$. It is the analogous for three particles of the in-medium two
particle G-matrix. The Bethe-Fadeev integral equation for $T^{(3)}$ is also
very similar to the integral equation for the G-matrix, where however in the
kernel the G-matrix appears in place of the bare NN interaction. This is in
line with the BBG expansion, where the bare NN interaction is systematically
replaced by the G-matrix in all the diagrams. In terms of $T^{(3)}$ the
contribution to the energy of the three hole-line diagrams can be depicted as
in Fig. \ref{fig:bhf3} (f), and it includes a direct and an exchange
contribution. For numerical convenience the diagrams with three G-matrices only
are usually separated from the diagrams with a larger number of G-matrices,
which will be indicated as ``higher order" three hole-line diagrams. The lower
order diagrams are depicted in Fig. \ref{fig:bhf3}, together with the
$U$-insertion diagram which contributes at this level of approximation. They
are indicated as ``bubble diagram" (c), $U$-insertion diagram (d) and ``ring
diagram" (e). Diagram (e) can be considered the exchange of diagram (c).
\par
The different three hole-line contributions to the interaction energy are
reported in Table I at different densities.

\begin{table}
\caption{Three hole-line contributions to the neutron matter EOS. $D_3$ is the
total three hole-line contribution, $B$ is the ``bubble diagram" of Fig. 1(c),
BU is the U-insertion diagram of Fig. 1(d), R is the ``ring diagram" of Fig.
1(e) and H indicates the ``higher order" diagrams, as defined in the text.
Energies are in MeV. }
\begin{ruledtabular}
\begin{tabular}{cccccc}

 $ k_{F}$ (fm$^{-1}$) & $D_3$ & $B$ & $BU$ &  $R$  & $H$ \\
\hline
  0.4 & 0.023&  -0.630  & 0.485 & 0.156 & 0.012  \\
  0.5 & 0.091&  -0.416  & 0.389 & 0.122 & -0.004  \\
  0.6 & 0.107&  -0.526  & 0.515 & 0.123 & -0.005  \\
  0.7 & 0.153&  -0.611  & 0.648 & 0.121 & -0.005  \\
  0.8 & 0.148&  -0.592  & 0.651 & 0.095 & -0.006  \\

\end{tabular}
\end{ruledtabular}
\end{table}

The overall three hole-line diagrams contribution $D_3$ is positive in this
density range and reaches a maximum around $k_F \approx 0.7$ fm$^{-1}$. This is
in line with the calculations at higher density, where $D_3$ turns actually
negative above a certain density. The absolute value of $D_3$ can be considered
small with respect to the two hole-line contribution $D_2$, but maybe it is not
completely negligible. In any case it is regularly decreasing with density and
at the lowest densities it becomes indeed negligible. As it is well known from
previous calculations, at densities higher than the ones considered here, the
smallness of $D_3$ is the result of a strong compensation among the different
contributions. While the ``higher order" terms, as defined above, can safely be
considered negligible, the absolute values of the bubble and $U$-insertion
diagrams, and, to a less extent, also the ring diagram are individually not
negligible, but their cancelation reduces by a large factor their overall
contribution. The smooth variation of each of these diagrams with density makes
the full three hole-line contribution decrease by almost one order of magnitude
from the highest to the lowest density.\par We take this result as an
indication of the convergence of the BBG hole-line expansion and we will assume
in the following that the total contribution of the diagrams with a number of
hole-lines larger than three can be neglected. To this respect it has to be
noticed that the bubble and U-insertion diagrams have opposite sign and their
absolute values must become closer and closer at lower density. In fact their
absolute values are not equal because the G-matrix of the bubble diagram is
``off shell", while the the G-matrix which defines the potential $U(k)$ of Eq.
\ref{eq:auxu}, as appearing in the $U$-insertion diagram, is  ``on shell". In
other words the entry energies of the two G-matrices, which are attached to the
particle line on the right in Fig. 1 (c) and in Fig. 1(d) , are different in
the two diagrams. However, it can be easily seen that the difference between
these two entry energies is a quantity proportional to the Fermi energy, and
therefore vanishing small at low enough density. The cancelation is therefore
expected and clearly apparent from the results of the explicit calculations. It
can be concluded that the three hole-line contribution further down in density
can be safely neglected.

\section{Discussion.}
\begin{figure}[]
\centerline{\includegraphics [height=80mm,width=120mm]{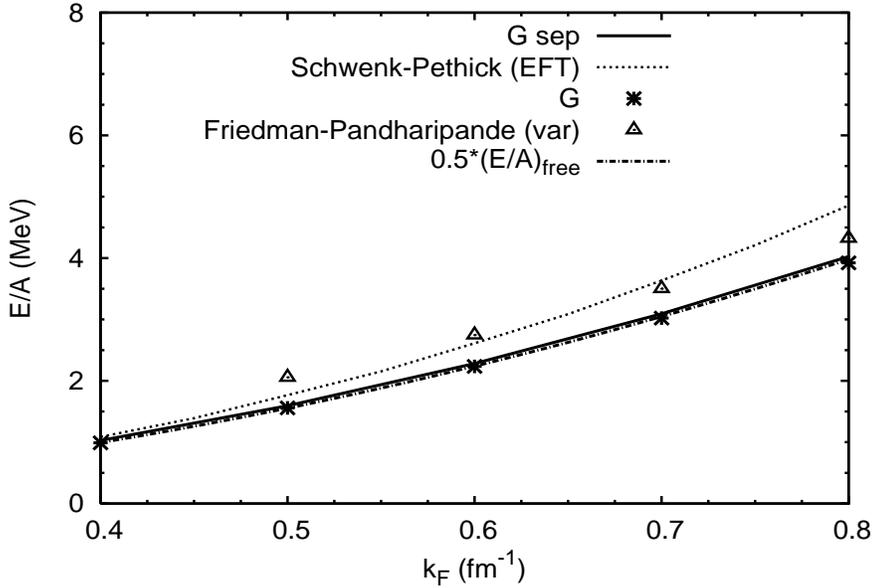}}
 \caption{Neutron matter EOS calculated within the BBG method (label G),
 within the variational method of ref. \cite{panda1} (triangles), according to the
 estimate of ref. \cite{Pethick} (dotted line) and with the separable representation of the
 G-matrix (label G sep). The dash-dotted line is one half of the
 free gas EOS}. \label{fig:eos}
\end{figure}

In the BBG expansion the EOS is given by the sum of the free kinetic energy and
the interaction energy discussed above. It is reported in Fig. \ref{fig:eos},
where we also show the neutron matter EOS estimated in ref. \cite{Pethick} and
the variational calculation of ref. \cite{panda1}. The contribution of the
three hole-line diagrams is hardly visible in the plot and it is neglected.

\par For comparison we also report the value of the free kinetic
energy divided by a factor 2. As already noticed in ref. \cite{panda2}, these
values stay surprisingly close to the full microscopic EOS. The agreement seems
to indicate that indeed the total energy is a function only of the Fermi
momentum $k_F$, as expected in the unitary limit. In the latter case, however,
Monte-Carlo calculations \cite{carl,boro,bulg1} suggest a factor 0.44 rather
than the value 0.5 found in our calculations. As noticed previously, the
unitary limit is not actually valid in neutron matter and the reason of such an
agreement must be found in some other more general considerations. To clarify
this point we take advantage on the above described results that the EOS is
determined to a good accuracy only by the G-matrix in the $^1S_0$ channel
calculated with the free single particle spectrum. We then construct a rank one
separable representation of the bare interaction which gives a free T-matrix,
also separable, with the known scattering length and effective range of the
neutron-neutron interaction in this channel and it actually depends only on
these two physical quantities. The corresponding in-medium G-matrix will be
also separable and will depend in addition on the Fermi momentum. If we take,
for simplicity, a Laurentzian form factor \beq \phi (k) \,=\,1 / (k^2 \,+\,
b^2) \label{eq:form} \eeq to be used for the bare neutron-neutron interaction,
the diagonal on-shell T-matrix can be written \beq \bra k | T | k \ket \,=\, a
/ \left[ 1 \,+\, u ( u + 2 -\beta )/(1 + \beta) \right] \le{sep} \eeq \noindent
where $ u = k^2/b^2 $ and the parameters $b$ and $\beta$ are determined by the
relationships \beq a \,=\, {1 \over b}\left( { {2\beta}\over {1 + \beta}}
\right)\ \ \ \ \ ; \ \ \ \ \ \
 r_0 \,=\, {1 \over b}\left( {{\beta - 2}\over {\beta}} \right)
 \le{para}
\eeq

\noindent

The values $a=-18.5$ fm and $r_0=2.7$ fm are used in the present calculation.
More details can be found in the Appendix. For large enough values of $b$, i.e.
small value of $r_0$, the separable representation is physically equivalent to
a zero range interaction with a smooth cut-off. However, the representation is
valid in the general case. In principle the effective range expansion holds at
small values of the momentum $k$, more precisely for $ k^2 << (ar_0)^{-1}$.
However, if the scattering is dominated by the virtual state of the
neutron-neutron $^1S_0$ channel, the separable representation can be valid in a
wider range of momenta. The function $\phi(k)$ can be then interpreted as the
form factor for the quasi-bound state (which is actually related to the
corresponding Gamow state \cite{gam,sep} ).\par
\begin{figure}[]
\centerline{\includegraphics [height=125mm,width=110mm]{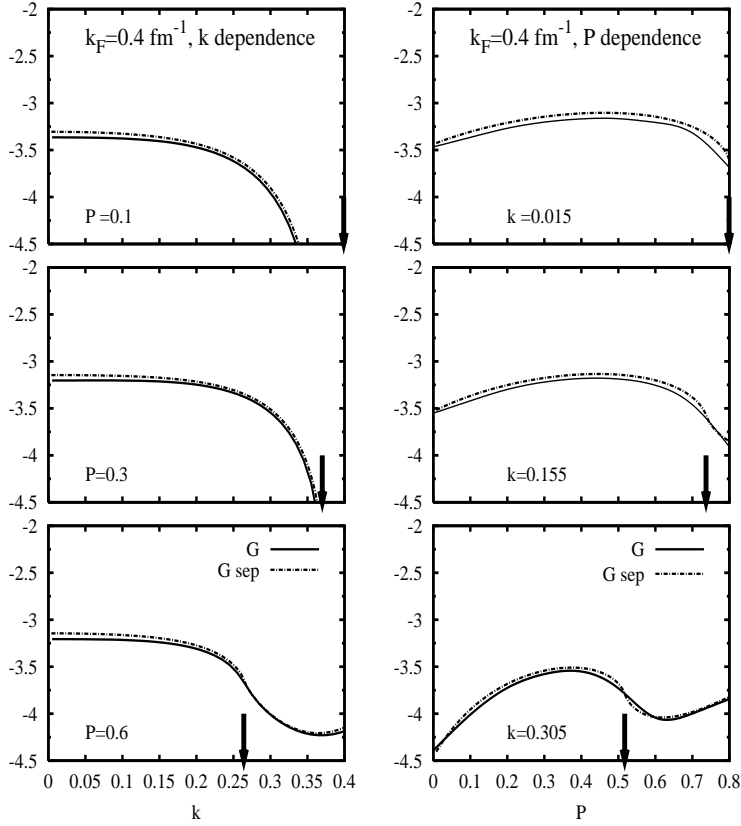}}
 \caption{Diagonal on shell G-matrix (full line) and the corresponding separable
 representation (dot-dashed line) at selected values of the total momentum P and
 relative momentum k and at the Fermi momentum $k_F = 0.4$fm$^{-1}$. The arrows
 have the same meaning as in Fig. \ref{fig:pauli}
 }
 \label{fig:sep1}
\end{figure}

\begin{figure}[]
\centerline{\includegraphics [height=125mm,width=110mm]{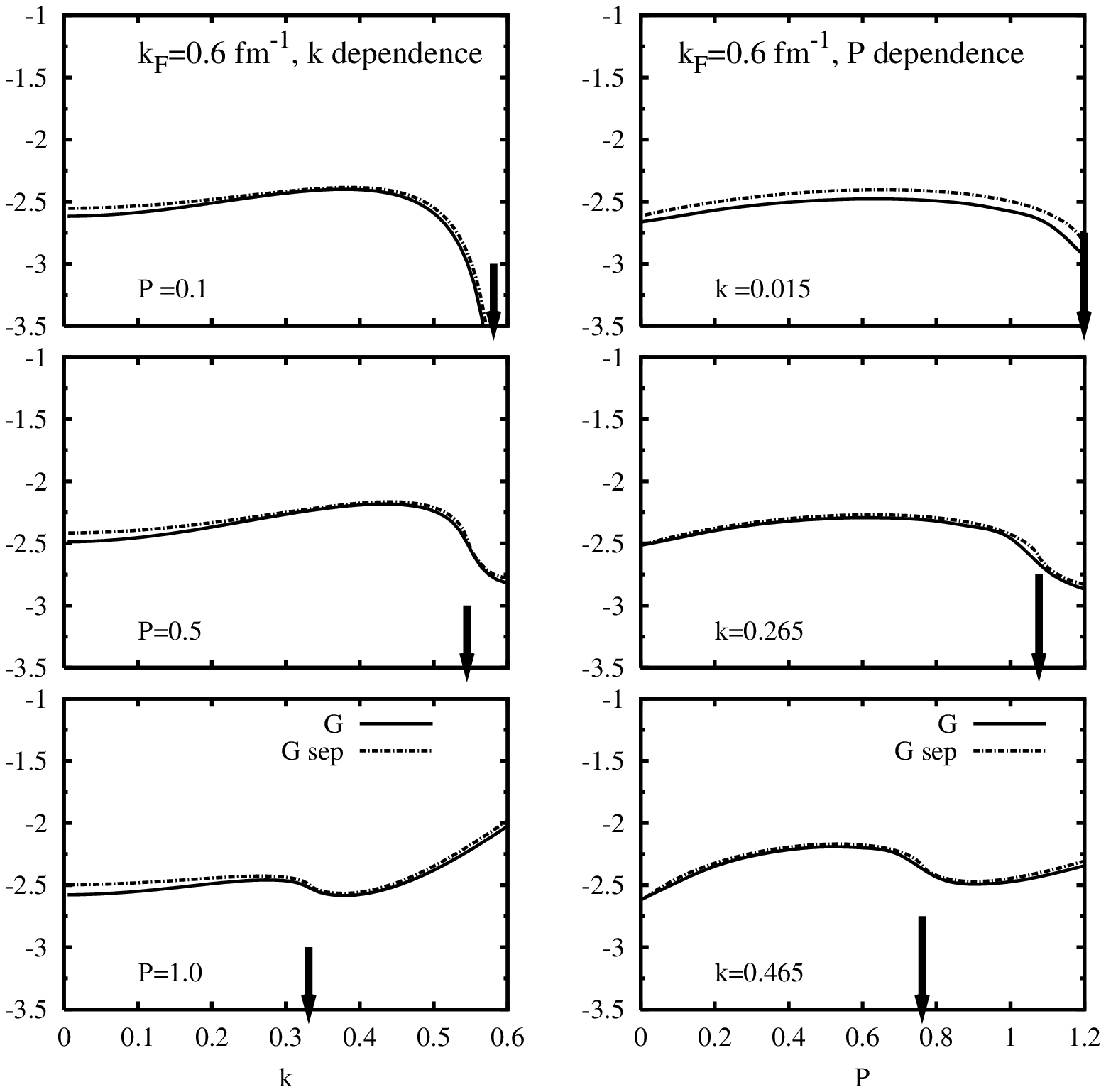}}
 \caption{Same as in Fig. \ref{fig:sep1}
 , but at $k_F = 0.6$fm$^{-1}$. }
 \label{fig:sep2}
\end{figure}

\begin{figure}[]
\centerline{\includegraphics [height=125mm,width=110mm]{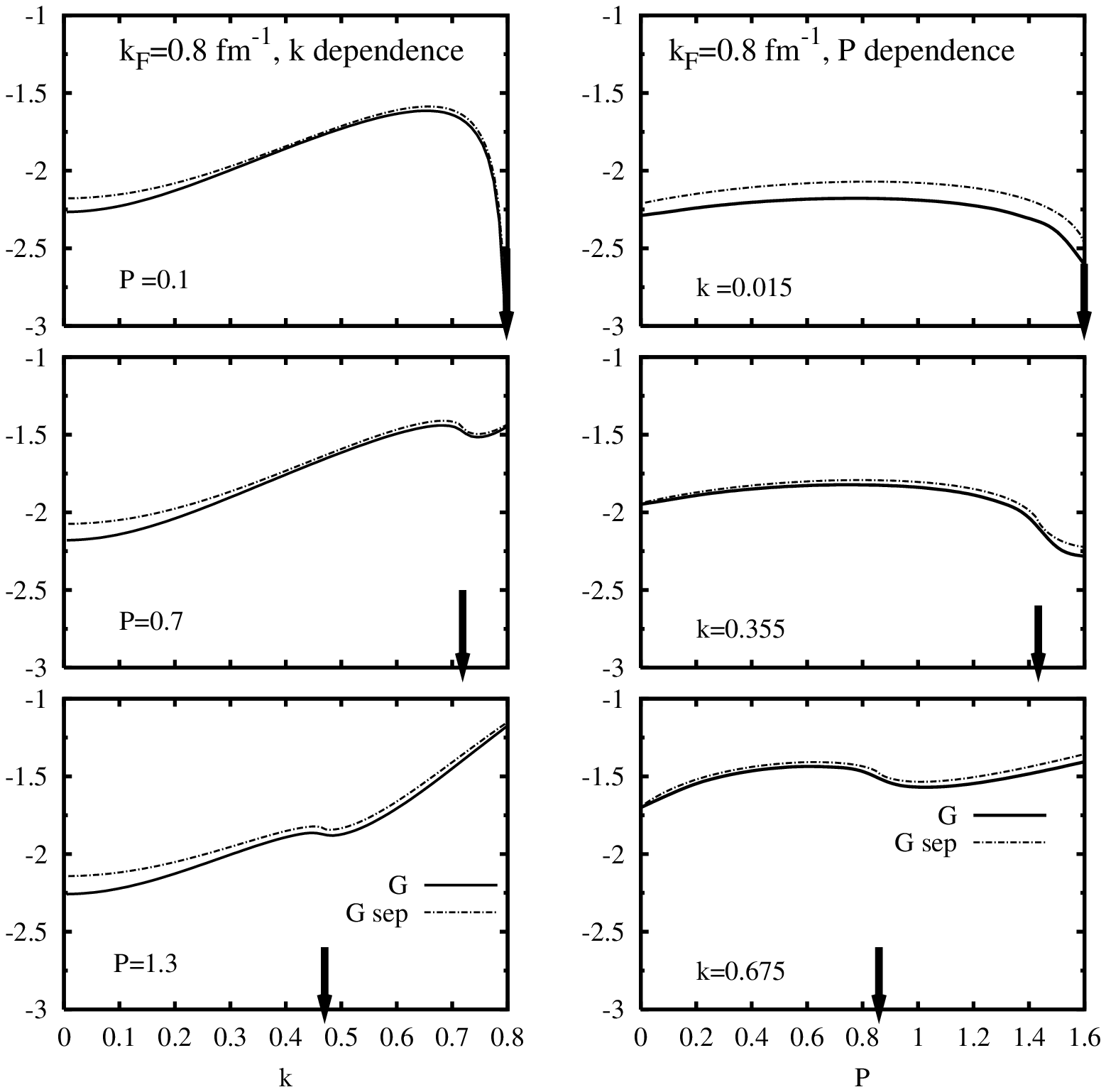}}
 \caption{Same as in Fig. \ref{fig:sep1}
 , but at $k_F = 0.8$fm$^{-1}$. }
 \label{fig:sep3}
\end{figure}

 The in-medium G-matrix
corresponding to the separable representation can be written as in Eq.
(\ref{eq:sep}), where however in the denominator an additional term appears,
whose explicit expression is given in the Appendix. The accuracy of the
separable representation can be appreciated in Figs.
\ref{fig:sep1},\ref{fig:sep2},\ref{fig:sep3}, where the exact (full line) and
separable (dashed line) diagonal on-shell G-matrices at different densities and
momenta are compared. The arrows have the same meaning as in Fig.
\ref{fig:pauli}.

\par
As expected, no virtual state is present for the G-matrix and correspondingly
the in-medium scattering length is strongly modified. As argued in ref.
\cite{panda2}, it becomes of order $k_F^{-1}$, while the effective range is
substantially reduced (see Appendix). Since the phase space present in the
calculation of the interaction energy is proportional to $k_F^3$, this can be a
possible explanation of the $k_F^2$ dependence of the binding energy per
particle \cite{panda2}, despite neutron matter is not strictly in the unitary
limit regime. The reduction of the total energy with respect to the free gas
value $E_{FG}$ by a factor so close to 2 is of course not easy to explain in
simple terms. The EOS calculated with the separable G-matrix is reported in
Fig. \ref{fig:eos}. It is indistinguishable with respect to the calculation
with the exact G-matrix. This shows that the bare interaction is mainly
determined by the virtual state (as embodied in the scattering length and
effective range values) and its in-medium suppression is the mechanism which
supersedes at the neutron matter EOS in the low density regime.\par In Fig.
\ref{fig:eos} we also report the EOS obtained within the variational method in
ref. \cite{panda1} as well as the result of the approximate estimate of ref.
\cite{Pethick}, based on effective theory methods. Some discrepancy is present
at the higher densities, but all the EOS seem to converge closely at the lower
densities, maybe with the variational results slightly apart. Actually the
variational calculation was performed with a different bare interaction, the
Urbana v$_{14}$ \cite{panda1}. However, we checked that the EOS calculated
within the BBG method with this different interaction is indistinguishable from
the one reported in Fig. \ref{fig:eos}. This is in line with the fact that the
interaction is determined solely by the scattering length and the effective
range, as the present analysis with the separable interaction clearly
indicates. The discrepancy is therefore not due to the different interactions
used but to the different adopted many-body methods. The variational results
are indeed slightly above the BBG results.
\par
\begin{figure}[]
\centerline{\includegraphics [height=80mm,width=120mm]{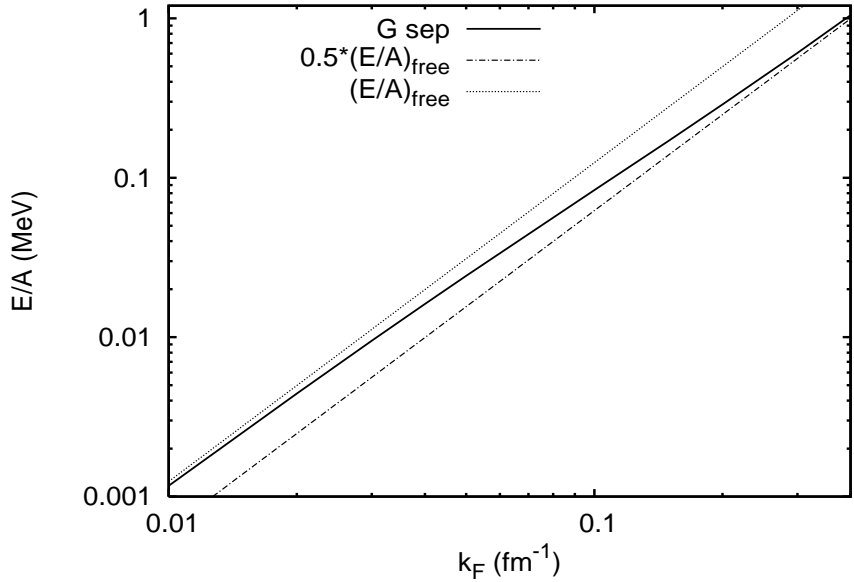}}
 \caption{Neutron matter EOS compared with the free gas one $E_{FG}$ and
 with $E_{FG}$/2 .} \label{fig:vlow}
\end{figure}

On the basis of the results of our analysis, it is possible to extend the
calculation of the EOS  below $k_F = 0.4$ fm$^{-1}$, just by using only the
separable G-matrix, since then the separable representation is even more
accurate. In Fig. \ref{fig:vlow} the EOS is reported in comparison with
${1\over 2} E_{FG}$.

One can see that deviations start now to appear, which indicates that the
simple rule of a factor $1/2$ is valid only in a limited range of density,
where indeed the neutron matter is ``closest" to the unitary limit. Indeed the
deviations start to appear for $k_F r_0  <  1$. Outside the considered range
$0.4 < k_F < 0.8$ fm$^{-1}$ the total energy is not ${1\over 2} E_{FG}$ and
even not proportional to $k_F^2$. At decreasing density the EOS is merging into
the free gas EOS, but this happens only at extremely low values, approximately
in agreement with the condition $k_F | a | << 1$. At these very low density
pairing could be of some relevance, but it cannot affect appreciably the total
energy since, once again, the unitary limit is not reached.

\section{Conclusions.}

In this paper we have established the pure neutron matter EOS on the basis of
the BBG many-body theory. We found that the EOS is determined with high
accuracy by the G-matrix in the $^1S_0$ channel only. In the range of density
corresponding to the Fermi momenta   $0.4 < k_F < 0.8$ fm$^{-1}$ the EOS turns
out to be very close to the value ${1\over 2}E_{FG}$, where $E_{FG}$ is the
free Fermi gas EOS. In this density range the condition $1/| a | < k_F/\alpha <
1/r_0$ is satisfied, with $\alpha = 2(9\pi/4)^{1\over 3}$ and $\alpha/k_F$ is
the average distance between particles. Here the scattering length $a$ and the
effective range $r_0$ are the ones of the $^1S_0$ neutron-neutron channel. We
interpret this results as an indication that neutron matter in this density
range is in some way ``close" to the unitary limit, even if such limit is
strictly not reached. In fact, for the unitary limit Monte Carlo calculations
\cite{carl,boro,bulg1} predict that the corresponding EOS should be
approximately $0.44 E_{FG}$. A rank one representation of the neutron-neutron
interaction, which is determined only by the scattering length and effective
range, proves to be extremely accurate in reproducing the G-matrix and the EOS.
Below $k_F = 0.4$ fm$^{-1}$ the EOS is not given by ${1\over 2}E_{FG}$ and in
the low density limit, when  $k_F | a | << 1$, it approaches smoothly the free
gas $E_{FG}$. Above $k_F = 0.8$ fm$^{-1}$ the EOS is not any more dominated by
the s-wave part of the NN interaction. As more complete calculations have shown
\cite{bal1,bal2}, in this case the EOS has no simple behavior.
\par\noindent
{\bf Acknowledgement}
\par\noindent
We gratefully thank Prof. P. Schuck for suggesting us the use of a rank one
approximation of the NN interaction suitable for the study of the neutron
matter EOS.

\par\noindent
{\bf Appendix}
\par
In this Appendix we give some details about the separable form of the
neutron-neutron interaction in the $^1S_0$ channel and the corresponding
in-medium G-matrix suitable in the low density region as discussed in the
paper. The rank-one representation of the interaction is written as \beq
 (k'|v|k)\, =\, \lambda\, \phi(k') \phi(k)
\label{eq:vsep} \eeq \noindent where the form factor $\phi(k)$ is given by Eq.
(\ref{eq:form}). The corresponding scattering T-matrix in free space can be
evaluated following the standard procedure for separable interactions
 \beq
 (k'|T(\omega)|k)\, = \lambda \phi(k') \phi(k) / [1\, -\,
 <\phi|G_0(\omega)|\phi> ]
\label{eq:tmat} \eeq

\noindent where $G_0$ is the free Green's  function and $\omega$ is the entry
energy.  With our choice of the form factor the integral for evaluating the
matrix element of the denominator can be performed analytically. The separable
interaction and the corresponding T-matrix depend only on the two parameters
$\lambda$ and $b$. Expanding the diagonal ($k' = k$) T-matrix for low momenta
and on-shell, i.e. at the kinetic energy $\omega = k^2/2\mu$, one can relate
these two parameters to the scattering length $a$ and effective range $r_0$ of
the original neutron-neutron interaction. This finally gives Eq.
(\ref{eq:para}) and the expression for the T-matrix in Eq. (\ref{eq:sep}).\par
Going to the in-medium G-matrix, one has simply to modify the integral for the
evaluation of the matrix elements of the free Green' s function by restricting
the integration above the Fermi surface. Still the integral can be done
analytically, and the final expression reads \beq
 (k|G(P,k_F)|k)\, =\, 1\,/\left[ \left( 1/a - {1 \over 2} r_0 k^2 + {1 \over 2}
 k^4/(b^3\beta) \right) + A(k,P,k_F) \right]
\label{eq:gmat} \eeq \noindent Neglecting the term $A(k,P,k_F)$ in the
denominator, one recovers the free T-matrix of Eq. (\ref{eq:sep}). The medium
effects are embodied in $A$, which reads
%\begin{equation}
\begin{eqnarray}
\!\!\!\!\!\!  A\,  = \,& - & {1 \over {\pi b}} (b^2 - k^2) \arctan \left({ {k_F
+ P/2}
\over b}\right) \nonumber \\
\!\!\!\!\!\!  & + & {1 \over {\pi }} k \log \left( {{k + k_F + P/2} \over {-k +
k_F + P/2}} \right)
\nonumber \\
\!\!\!\!\!\! & + & {1 \over {\pi P}}(k_F^2 - P^2/4 - k^2) \log \left\vert {{
(k_F + P/2)^2 + b^2 } \over {k_F^2 - P^2/4 + b^2}}\cdot { {k_F^2 - P^2/4 - k^2}
\over { (k_F + P/2)^2 -k^2 } } \right\vert \label{eq:A}
\end{eqnarray}
% \eeq
 \noindent where $P$ is the total momentum of the
two particles. Expanding the G-matrix at small momentum and at $P = 0$, one can
obtain the in-medium scattering length $a'$ and effective range $r_0'$. They
read
\begin{eqnarray}
 a' \, & = & \, a / ( 1 -  { {a b} \over \pi } \arctan ( {k_F \over b} ) ) \approx
-{\pi\over {2k_F}}  ;  \\
r_0'\,  & \approx &\, r_0\, -\, {4\over {\pi k_F}}
 \label{eq:ar}
 \end{eqnarray}
\noindent The expression for $r_0'$ is obtained assuming $ k << k_F $, so it
cannot be considered at too low density.


\begin{thebibliography}{99}


\bibitem{shap} S.L. Shapiro and S.A. Teukolsky, {\it Black Holes,
White Dwarfs and Neutron Stars} (Jhon Wiley and Sons, New York, 1983).
\bibitem{carl} J. Carlson, S.-Y. Chang, V.R. Pandharipande andK.E. Schmidt,
Phys. Rev. Lett. 91, 050401 (2003).
\bibitem{boro} G.E. Astrakharchik, J. Boronat, J. Casulleras and S. Giorgini,
Phys. Rev. Lett. 93, 200404 (2004).
\bibitem{bulg1} A. Bulgac, J.E. Drut and P. Magierski,
Phys. Rev. Lett. 96, 090404 (2006).
\bibitem{bulg2} A. Bulgac, J.E. Drut and P. Magierski,
Phys. Rev. Lett. 99, 120401 (2007).
\bibitem{panda1} B. Friedman and V.R. Pandharipande, Nucl. Phys. 361 A, 501
(1981).
\bibitem{panda2} J. Carlson, J. Morales, V.R. Pandharipande and D.G. Ravenhall,
Phys. Rev. C 68, 025802 (2003).
\bibitem{Pethick} A. Schwenk and C.J. Pethick, Phys. Rev. Lett. 95, 160401
(2005).
\bibitem{book} For a pedagocical introduction see
{\it  Nuclear Methods and the Nuclear Equation of State}, M. Baldo Ed.,1999,
World Scientific, Singapore.
\bibitem{wir} Wiringa R B, Stoks V G J and Schiavilla R 1995, Phys.
Rev. {\bf C 51}, 38.
\bibitem{bbb} M. Baldo, I. Bombaci and G.F. Burgio, Astronomy and
Astrophysics 328 (1997) 274; X.R. Zhou, G.F. Burgio, U. Lombardo,
H.-J. Schulze and W. Zuo, Phys. Rev. C 69 (2004) 018801.
\bibitem{jpg} M. Baldo and C. Maieron, J. Physics G:Nucl. Part. Physics 34,
R243 (2007).
\bibitem{bal1} M. Baldo, C. Maieron, P. Schuck and X. Vinas, Nucl. Phys. A 736,
241 (2004).
\bibitem{bal2} M. Baldo, A. Fiasconaro, G. Giansiracusa, U. Lombardo and H.Q.
Song, Phys. Rev. C 65, 017303 (2001).
\bibitem{bethe} R. Rajaraman R and  H. A. Bethe, Rev. Mod. Phys.  39, 745 (1967).
\bibitem{gam} G. Gamow, Z. Phys. 51, 204 (1928).
\bibitem{sep} M. Baldo, L.S. Ferreira and L. Streit, Phys. Rev. C 32, 685
(1985).










\end{thebibliography}
\end{document}